\documentclass[preprint,review,sort&compress]{elsarticle}

\biboptions{numbers,compress}
\usepackage{amssymb}
\usepackage{graphicx}
\usepackage{tabularx}
\usepackage{float}
\usepackage{subcaption}
\usepackage{url}
\usepackage[all]{nowidow}
\usepackage[percent]{overpic}
\usepackage[shortcuts]{extdash}

\makeatletter
\providecommand{\doi}[1]{%
  \begingroup
  \let\bibinfo\@secondoftwo
  \urlstyle{rm}%
  \href{http://dx.doi.org/#1}{%
    doi:\discretionary{}{}{}%
    \nolinkurl{#1}%
  }%
  \endgroup
}
\makeatother

\graphicspath{ {./} }

\journal{Nuclear Instruments and Methods in Physics Research Section A}

\begin{document}

\begin{frontmatter}

  \title{In-Flight Performance of the Advanced Radiation Detector for UAV
    Operations (ARDUO)}
  \author[cu,chis]{C.M.~Chen}
  \author[cu,chis]{L.E.~Sinclair\corref{cor1}}
  \ead{laurel.sinclair@canada.ca}
  \author[gsc]{R.~Fortin}
  \author[gsc]{M.~Coyle}
  \author[cu,ets]{C.~Samson}
  \cortext[cor1]{Corresponding author}
  \address[cu]{Department of Earth Sciences, Carleton University, Ottawa, Ontario, Canada}
  \address[chis]{Canadian Hazard Information Service, Natural Resources Canada,
    Ottawa, Ontario, Canada}
  \address[gsc]{Geological Survey of Canada, Natural Resources Canada, Ottawa, Ontario, Canada}
  \address[ets]{Department of Construction Engineering, \'Ecole de Technologie
    Sup\'erieure, Montr\'eal, Qu\'ebec, Canada}

  \begin{abstract}
      Natural Resources Canada is responsible for the provision of aerial
      radiometric surveys in the event of a radiological or nuclear emergency in
      Canada. Manned aerial surveys are an essential element of the planned
      consequence management operation, as demonstrated by the recovery work
      following the 2011 Tohoku earthquake and tsunami, and their effects in
      Fukushima, Japan. Flying lower and slower than manned aircraft, an unmanned
      aerial vehicle (UAV) can provide improved spatial resolution.  In
      particular, hot spot activity can be underestimated in manned survey results
      as the higher flight altitude and wider line spacing effectively average the
      hot spot over a larger area. Moreover, a UAV can enter an area which is too
      hazardous for humans, due not only to the radiological threat which is its
      target, but also to other anticipated hazards such as explosives,
      airborne chemical hazards, or open water.
      Natural Resources Canada has been investigating the
      inclusion of UAV-borne radiation survey spectrometers into its aerial survey
      response procedures. The Advanced Radiation Detector for UAV Operations
      (ARDUO) was developed to exploit the flight and lift capabilities available in
      the under 25~kg class of UAVs. The detector features eight 2.8~cm~$\times$~2.8~cm~$\times$~5.6~cm
      CsI(Tl) crystals arranged in a self-shielding configuration, read out with
      silicon photomultipliers. The signal is digitized using miniaturized custom electronics.
      The ARDUO is flown on a main- and tail-rotor UAV called Responder which has a 6~kg lift
      capacity and up to 40~minute endurance. Experiments were conducted to
      characterize the performance of the ARDUO and Responder UAV system in both
      laboratory and outdoor trials. Outdoor trials consisted of aerial
      surveys over sealed point sources and over a distributed source of 10~MBq/m$^2$ of
      La\-/140. Results show how the directional response of the
      ARDUO can provide an indication in real time of source location to guide the
      UAV during flight.
      As well, the results show how utilization of the directional
      information in post-acquisition processing can result in improved
      spatial resolution of radiation features for both point and distributed sources.
  \end{abstract}


\end{frontmatter}


  \section{Introduction}
  Natural Resources Canada (NRCan) is
  responsible for conducting radiometric surveys in response to radiological
  and nuclear incidents. This requires high sensitivity systems that
  are easily deployable in order to detect, locate and identify radioactive
  materials in real time. Manned aircraft and terrestrial vehicle systems are
  traditionally used for nuclear emergency response surveying, and have demonstrated
  their importance during the Fukushima Dai-ichi Nuclear Power Plant accident \cite{iaea2003, sinclair2011, sanada2014, tanigaki2013}.
  As the use of unmanned aerial vehicles (UAVs) becomes more common, NRCan is
  investigating the benefits UAVs provide to their existing radiometric
  systems. Flying lower and slower than manned aircraft, a UAV can provide improved spatial resolution. This is particularly beneficial in
  hot spot activity assessments as lower resolution surveys can underestimate
  count rates by averaging the hot spot over a larger area~\cite{sinclair2018}.
  UAVs are also able to provide an additional level of safety for
  operators as they are able to collect measurements in areas containing
  possible radiological threats, or associated threats such
  as explosive or chemical hazards.\par Several radiation detecting UAV systems have been built and tested
  \cite{sanada2015, kochersberger2014, kurvinen2005, li2018, pollanen2009, macfarlane2014}. Heavier systems have larger payload capacities,
  but have limited use due to Canadian flight regulations. Smaller UAVs are
  only capable of carrying smaller radiation detectors that have fewer
  applications for
  nuclear emergency response situations. 
  The radiation detecting UAV system designed at NRCan has been specifically
  optimized for performance in the under 25~kg class of UAVs set by Canadian
  regulations.

  Using the EGSnrc~\cite{egs1,egs2} radiation propagation package, NRCan 
  designed the Advanced Radiation Detector for UAV Operations (ARDUO)~\cite{arduoSinclair}
  to have optimal sensitivity and direction reconstruction capability
  given the lift capacity of the ING Robotic Aviation Inc.\ main- and
  tail-rotor Responder UAV~\cite{ing}.
  The ARDUO features thallium-doped cesium iodide (CsI(Tl)) scintillator
  detectors in a self-shielding configuration read out with silicon photomultipliers.  The performance of the
  silicon photomultipliers had been  characterized though previous research~\cite{imager}.
  Radiation Solutions Inc. \cite{rsi} developed custom miniaturized pulse
  shaping and digital electronics for ARDUO and as well designed the ARDUO
  housing and mounts and integrated ARDUO's direction calculation into their
  RadAssist data acquisition and real-time display software.
  Post-acquisition processing
  techniques use this direction information to improve source
  localization. 
  This paper presents results on how the ARDUO performs in a series of
  laboratory tests, a
  small-scale point-source survey and a larger-scale distributed-source 
  survey.
  Further details about this work can be found in~\cite{Chen_2018}.

  \section{Instrumentation}
\label{instrumentation}
  The ARDUO consists of eight 2.8~cm~$\times$~2.8~cm~$\times$~5.6~cm CsI(Tl) crystals in a
  self-shielding arrangement, resulting in a sensitive volume of 0.35~L. Each crystal is coupled with a silicon
  photomultiplier for light collection and the signal is digitized with
  miniaturized custom electronics. Every second, eight 1024-channel energy
  spectra spanning the energy range up to 3~MeV
  (one spectrum from each
  crystal) are saved and are tagged with a position from the
  Global Navigation Satellite System (GNSS). Each crystal in the ARDUO has a
  full-width-at-half-maximum energy
  resolution of approximately 7~\% at 662~keV.

ARDUO is mounted on the Responder UAV
  built by ING Robotics Inc. The Responder is a main- and tail-rotor, battery-powered UAV
  with a lift capacity of up to
  6~kg and an endurance of up to 40~minutes. This type of UAV was chosen over
  a fixed-wing aircraft as it is capable of vertical takeoffs and landings, and has the ability
  to hover. In addition, main- and tail-rotor UAVs have larger lift capactities and better stability in windy conditions when compared to similar-sized
  multi-rotor UAVs.\par
\begin{figure}[!b]
    \centering
    \includegraphics[width=0.7\columnwidth]{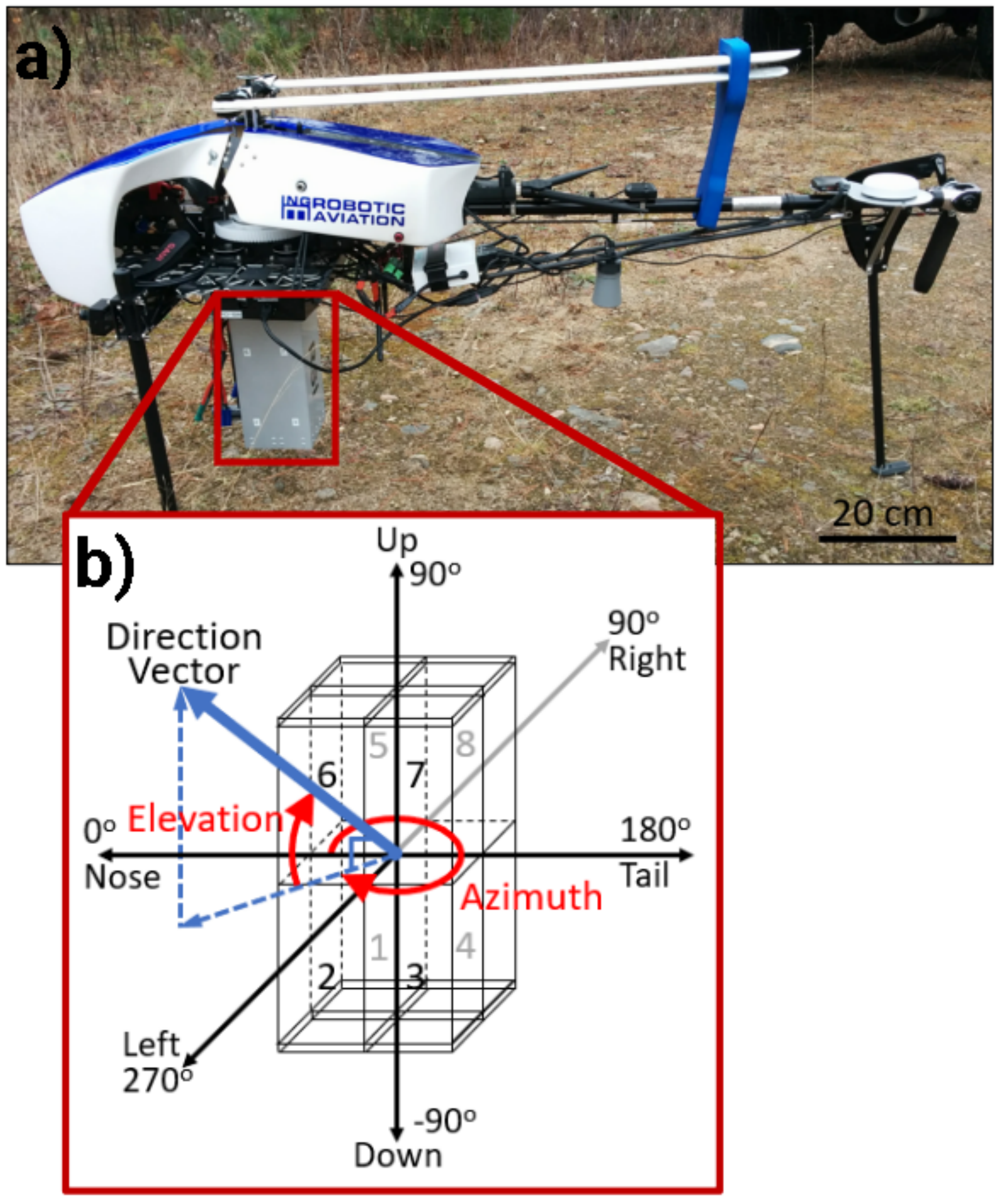}
    \caption{a)~ING Responder UAV with the ARDUO mounted underneath. b)~The ARDUO
      crystal arrangement and direction vector orientation. Individual crystals
      are numbered from 1 to 8. The nose and tail directions are
      indicated. The elevation and azimuth angles for a representative
      reconstructed direction vector are shown.}
    \label{fig:uavarduo}
  \end{figure}

The Responder UAV and ARDUO crystal layout are shown in
Figure~\ref{fig:uavarduo}.
The ARDUO is mounted below the center of mass of the
Responder UAV. The crystals are arranged in a rectangular array with two
layers of four crystals. A
representative direction vector is shown, as well as its associated azimuth
and elevation angles. The ARDUO is oriented such that the 0$^\circ$ azimuth
axis is aligned with the nose of the Responder UAV.\par
  The ARDUO and Responder UAV system is flown in a
  semi-autonomous mode where it follows a series of GNSS waypoints that are planned by
  the operator prior to flight. During flight, the Responder UAV's PixHawk~1
  autopilot logs the position and orientation of the system. The autopilot
  uses an extended Kalman filter with measurements from a GNSS antenna and receiver, compass, barometer
  and inertial measurement unit to determine its position and orientation
  every 0.02 seconds. The position and orientation data logged by the the
  autopilot are available for post-acquisition data analysis and mapping.

  \section{Direction Reconstruction}
  \label{dirveccalc}
  The direction vector pointing towards a detected radiation source is calculated using the self-shielding configuration of
  the crystals. The crystals with relatively higher gamma count
  rates will be closer to the source, while the crystals behind will have
  relatively lower gamma count rates due to the shielding of the front crystals. Using
  the relative count rates in each of the eight crystals over an energy range of 0.1 to 3.0~MeV, a direction vector in
  three dimensions, $\textbf{v} = (u,v,w)$, is calculated according to Equations
  \ref{eq:dirvec1} to \ref{eq:dirvec2}, where $c_i$
  is the number of counts per second in crystal number $i$. 
  \begin{equation}
    u = c_1 + c_5 - c_3 - c_7.
    \label{eq:dirvec1}
  \end{equation}
  \begin{equation}
    v = c_2 + c_6 - c_4 - c_8.
  \end{equation}
  \begin{equation}
    w = c_5 + c_6 + c_7 + c_8 - c_1 - c_2 - c_3 - c_4.
    \label{eq:dirvec2}
  \end{equation}
The direction vector is calculated over this
wide range of energy deposits (0.1~MeV to 3.0~MeV) to allow the direction reconstruction algorithm to produce a
result for any spectrum in real time, prior to isotope identification.
The direction reconstruction algorithm can also be expected to return a sensible direction
result for heavily shielded isotopes or mixed radiation fields due to this
wide energy window range.

After calculating the direction vector in three dimensions, the vector is
comverted into spherical polar coordinates using Equations~\ref{eq:phi}~and~\ref{eq:theta}.
  \begin{equation}
    \mathrm{Azimuth} = \mathrm{arctan}\left(\frac{v}{u}\right) - 45^\circ,
    \label{eq:phi}
  \end{equation}
and,
  \begin{equation}
   \mathrm{Elevation} = \mathrm{arccos}\left(\frac{w}{\sqrt{u^2 + v^2 + w^2}}\right) - 90^\circ.
    \label{eq:theta}
  \end{equation}

The elevation angle spans from -90$^{\circ}$ (directly below ARDUO)
to 90$^{\circ}$ (directly above ARDUO), and the azimuth angle spans
360$^{\circ}$ clockwise along the horizon from the nose direction of the UAV. These angles are shown in Figure~\ref{fig:uavarduo}b.\par
\begin{figure}[!htb]
    \centering
    \includegraphics[width=\textwidth]{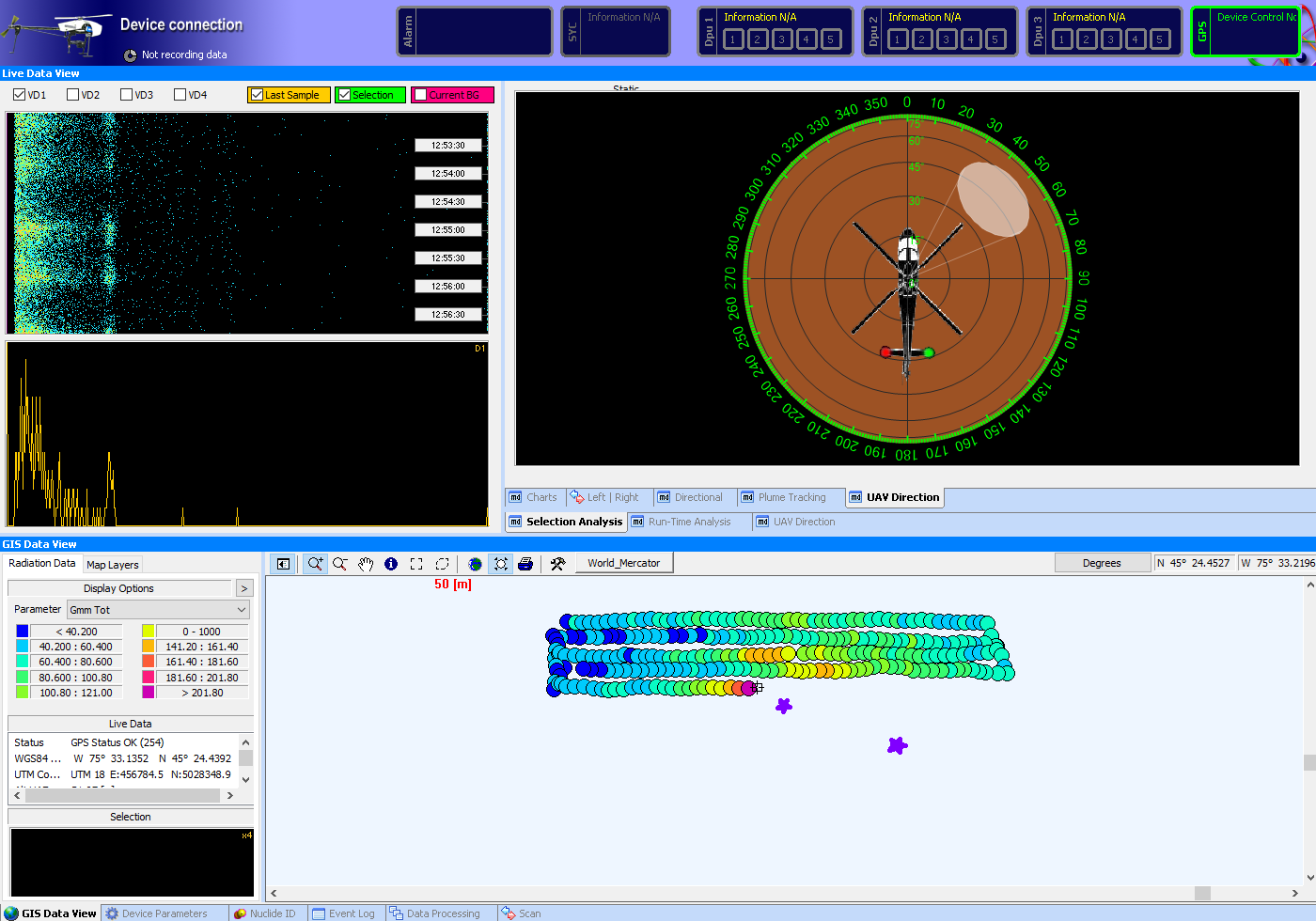}
    \caption{RadAssist~\cite{rsi} real-time data view during an aerial survey over two $^137$Cs sources. 
	The top left screen shows time on the y\-/axis, energy on the x\-/axis, and counts on the z\-/axis shown by the colour scale. 
	The middle left screen shows the current one-second energy spectrum for one detector. The energy scale in both these screens is a linear 
	domain from 0 to 3.072~MeV. The top right screen shows the current direction reconstruction from one second of data indicating a source 
	to the front and right of the aircraft's direction of propagation at an elevation angle of about 45$^\circ$. The white circle indicates 
	that the elevation is downward. Upward or positive elevation angle reconstructions would be indicated with a blue circle. The bottom 
	screen shows the GNSS location of each data point on a colour scale from blue (low count rate) to purple (high count rate). Here 
	purple stars have been overlaid to show the true locations of the two sources. Unless the source location is known a priori, this is 
	not usually available in real time. The real-time view is refreshed every second.}
    \label{fig:realtimedisplay}
  \end{figure}

  Incorporating the telemetry of the Responder UAV, the ARDUO control
  software displays the self-shielding direction vector to the operator
  in real time. The display is
  updated every second when a new radiation data point is collected. A screenshot of the real-time display is shown in
  Figure~\ref{fig:realtimedisplay}.
This display shows the counts in
  the current energy spectrum versus time; the one-second spectrum; a map of each data
  point location coloured according to the total spectrum counts; as well as
  the directional information. In the
  screenshot provided, it can be seen in the directional display that the
  source is located towards the right with respect to the flight direction, at an approximate -45$^\circ$
  angle downwards. This corresponds well to the true locations of the sources indicated by purple stars in the map display.  
  The operator can also use the energy
  spectrum information displayed in real time for isotope identification.

  \section{Post-Acquisition Data Analysis and Mapping}
  \label{processing}
  After a survey is flown, the data is processed and mapped to examine the
  spatial distribution of any sources contained within the survey area. Processing follows a number of distinct
  steps, as explained in the following subsections.

  \subsection{Energy Window Count Rates}
  A spectrum energy window from 0.1~to~3.0~MeV is used to calculate the
  crystal count rates per data point for the direction reconstruction process described in
  Section~\ref{dirveccalc}. The spectrum is then examined to determine the
  presence of any photopeaks and for isotope identification. An
  isotope-dependent energy window is then established around the isotope's
  photopeak of interest. The ARDUO count rates in this isotope-dependent energy window are used for subsequent count-rate mapping. 

  \subsection{Background Subtraction}
  Aerial surveys are generally conducted outdoors, therefore the
  naturally-occurring isotopes of K\-/40, U\-/238, and Th\-/232 could contribute a significant
  background signal to each measurement. This background signal can be
  estimated within the survey area by conducting a nearby survey where only natural sources
  are present; or, if available, by using previously conducted radiometric surveys for
  geological purposes of the area or nearby surroundings. The average background
  count rate is subtracted from the data.\par

  \subsection{Lag Correction}
  A lag correction is applied to the ARDUO radiation data time-stamps
  to account for the ARDUO's motion during the one-second data accumulation time. As each data point is time-stamped at the
  beginning of the accumulation time, 0.5~seconds is added to each
  time-stamp to shift them to a midpoint time. These time-stamps are used to
  correlate the ARDUO radiation data with the Responder UAV's autopilot data.

  \subsection{Location and Orientation}
  The Responder UAV's PixHawk~1 autopilot logs its latitude,
  longitude, altitude above ground level, and orientation every 0.02~seconds using an extended
  Kalman filter and the suite of sensors listed in
  Section~\ref{instrumentation}. This data stream is merged with the ARDUO's
  radiation data stream using the lag-corrected time-stamps, thus providing each
  radiation measurement with a position and orientation for further analysis. 

  \subsection{Heading Correction}
  \label{headingcorr}
  As the Responder and ARDUO UAV system is flying, the flight direction
  makes some angle with respect to true north. The nose of the 
  Responder,
  which is in-line with the ARDUO's 0$^\circ$ azimuth angle axis, can
  be rotated from this flight direction due to windy conditions or
  during turns. These rotations are logged by the autopilot as the yaw, the
  orientation angle of the nose of the Responder
  with respect to true north. The direction vector's azimuth angle is calculated with respect to
  the yaw angle of the Responder at the time the data point is collected. The
  direction vector's azimuth angle with respect to the ARDUO is added to the
  yaw angle of the Responder at the time the data point is collected to 
  obtain an azimuth angle with respect
  to true north. 

  The pitch and roll angles would also have an effect on the direction
  vectors.  Inspection of the flight controller records from one survey
  verified that the average pitch and roll angles, including turns where the
  roll and pitch are more variable,  
  were $0^\circ \pm 2^\circ$ and $4^\circ \pm 3^\circ$
  respectively, where the uncertainty shows the root-mean-square deviation from
  the mean value.  
  These small deviations of the system's orientation from a level orientation (with
  the ARDUO $w$ axis pointing upward) were neglected in the analysis.

  \subsection{Count-Rate Map}
  With all preceding corrections made, the latitude and longitude from the
  autopilot are
  converted from the geographic World Geodetic System 1984 system to
  easting and northing coordinates from the projected Universal Transverse
  Mercator system in the corresponding zone of the survey. Each data point is mapped based on
  its location. The count rate in the
  isotope-dependent energy window for each data point is interpolated using
  the inverse distance interpolation algorithm provided by the ArcGIS
  software suite \cite{arcgisIDW}. The
  interpolated map can be used to interpret the spatial distribution of the
  radiation in the survey area.
  
  \subsection{Direction Vector Overlay}
  The direction vectors are calculated using the count rates in the spectrum energy
  window (0.1~to~3.0~MeV) and the process described in
  Section~\ref{dirveccalc}. This energy window is used for the direction
  vector calculation as the partial
  gamma-ray energy deposits are part of the signal from the source, therefore
  increasing the signal to noise ratio. All preceding
  corrections are made plus the additional heading correction to the azimuth
  angle presented in Section~\ref{headingcorr}. The direction vectors can be displayed in
  two-dimensions and overlaid on the count-rate map. Each direction vector is
  displayed as an arrow with the arrowhead at the data point location. The
  arrow is rotated to point in the direction of the azimuth angle with
  respect to true north. The length of the arrow represents the elevation
  angle, with shorter arrows being closer to vertical, and longer arrows
  being closer to horizontal. The arrow colour indicates whether the elevation
  angle is positive (upwards), or negative (downwards). 

  \subsection{Direction Vector Projection}    
  An additional processing step using the direction vectors can be done to
  improve on source localization. Each direction vector is projected to the ground
  surface, and new point is created where the vector intersects the ground. Only data
  points with downward elevation angles are used in this part of the
  analysis.  An additional restriction to elevation angles less than
  $-40^\circ$ was applied to avoid the horizon region where reconstruction of
  the elevation angle is poor (to be presented in Section~\ref{subsec:labtests}).
  \par
\begin{figure}[!htb]
    \centering
    \includegraphics[width=\columnwidth]{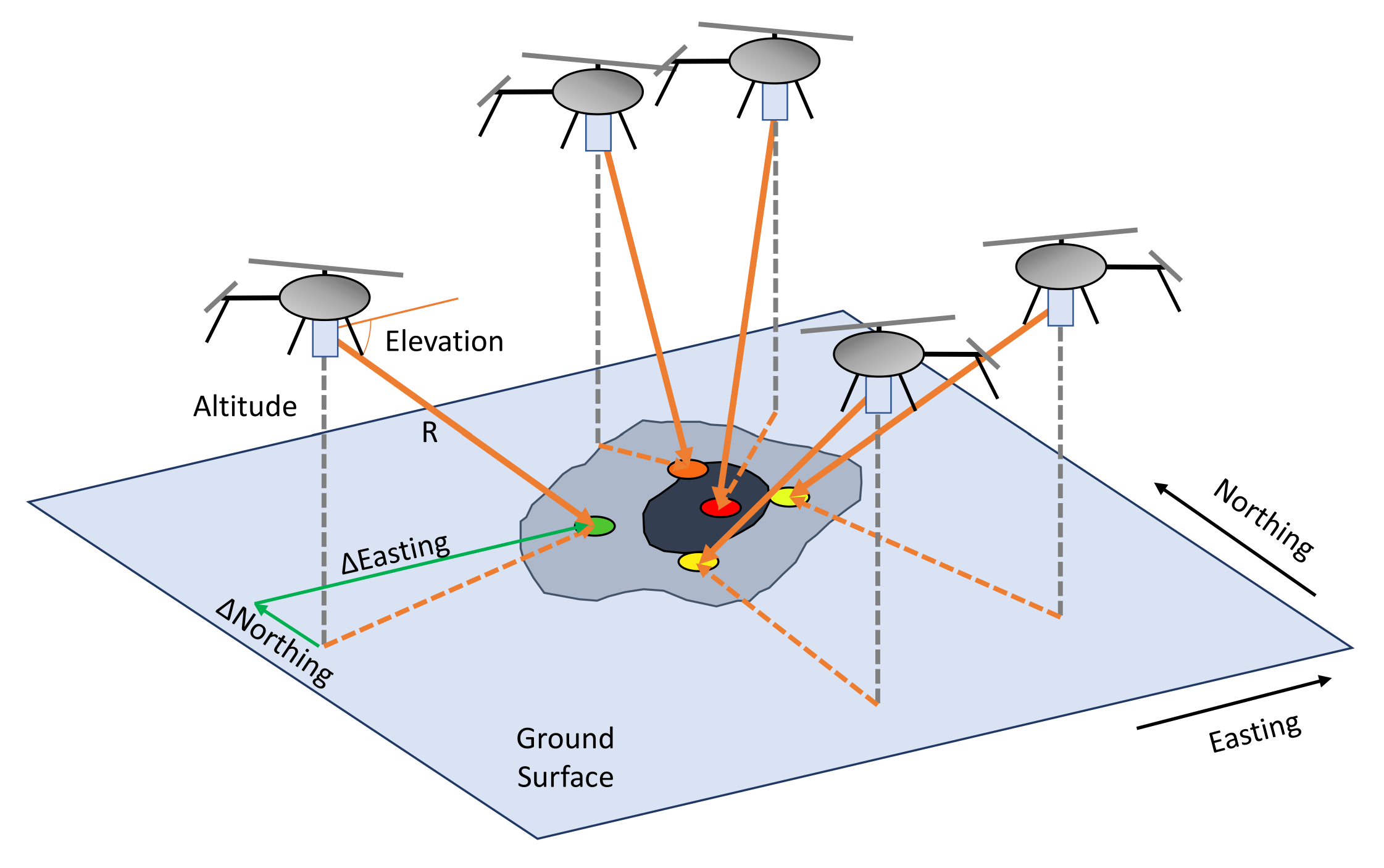}
    \caption{Visualization of the process of projecting the direction vectors
      collected during flight down to a ground surface. The ground surface is
      represented by the blue plane. The direction vector is represented by
      the orange arrow that is projected from the ARDUO under the helicopter
      to the ground surface. The coloured points are the locations where the
      direction vectors intersect the ground. They are coloured based on the
      original data point's count rate, with red indicating a higher count
      rate and green indicating a lower count rate. The darker blue contours on the
      ground surface represent the spatial point density contour results from
      the ArcGis \cite{arcgisKD} kernel density analysis. The middle contour
      with the darkest blue area indicates the most likely source location.}
    \label{fig:projschem}
  \end{figure}

  A diagram illustrating this process is shown in Figure~\ref{fig:projschem}.
  The
  locations of each data point at an altitude above ground level are shown by
  the helicopters. The direction vector is shown by the orange arrow extending
  from the ARDUO to the ground, shown by the blue surface. While a flat
  surface is shown in Figure~\ref{fig:projschem}, the method is
  extendable to terrain with topographic variations by using a digital elevation model. The direction vector's intersection point
  with the ground is calculated using the Responder
  UAV's orientation and position from the autopilot, and the direction
  vector's azimuth and elevation angles. The
  kernel density function provided by ArcGIS \cite{arcgisKD} is then used to
  contour the spatial point density of the intersection points, where each
  point is weighted by the number of measured gamma-ray energy deposits.  
  Areas with higher point densities indicate possible source locations.\par
  As the distance between the location where the data was collected and the location where the direction vector intersects the ground
  is calculated, a source strength at ground level could in principle be
  estimated. This estimation adds additional complexity to the processing methods,
  especially in the case of multiple point sources, and is outside the scope
  of this paper. 

  \section{Results}

  \subsection{Laboratory Tests}
  \label{subsec:labtests}
\begin{figure}[!htb]
    \centering
    \begin{subfigure}[t]{0.5\columnwidth}
      \centering
      \includegraphics[width=\columnwidth]{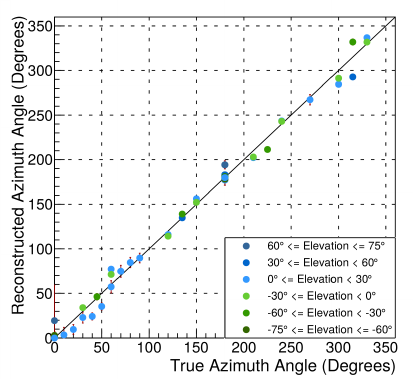}
\caption{Reconstructed azimuth angles.}
      \label{fig:az}
    \end{subfigure}
    \begin{subfigure}[t]{0.5\columnwidth}
      \centering
      \includegraphics[width=\columnwidth]{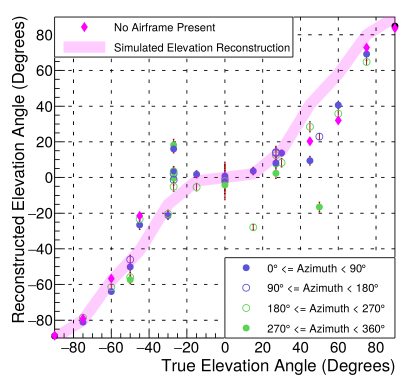}
      \caption{Reconstructed elevation angles.}
      \label{fig:el}
    \end{subfigure}
    \caption{Results from the laboratory direction
      reconstruction tests. Each data point represents the mean value of the
      reconstructed direction vector angles from 30 ten-second datasets. The error
        bars show the root mean square deviation of the ten-second
        measurements. a)~shows the reconstructed azimuth angle versus true azimuth angle for a variety of
        elevation angles. All measurements were taken with the airframe
        present. b)~shows the reconstructed elevation angle versus true elevation angle for a
        variety of azimuth angles. The pink shaded curve is an approximation of the known
        system response from Monte Carlo studies. The pink diamonds are measurements taken without the
        airframe.}
    \label{fig:dirrec}
  \end{figure}

  To determine the the ARDUO's direction reconstruction capabilities, a series
  of experiments were conducted in laboratory conditions. A calibrated 3.2~MBq point
  source of Cs\-/137 was placed at known
  locations around the ARDUO mounted on the Responder UAV. Locations were chosen at
  varying azimuth and elevation angles to cover a range of three-dimensional
  space. Distances from the source to the ARDUO ranged from approximately 1~m
  to 3~m. A
  subset of these locations was used for measurements without the UAV
  airframe present to determine its influence. A five-minute measurement was taken at each source location and split
into an ensemble of 30~ten\-/second datasets. Background measurements, lasting
fifteen minutes, were collected daily and used to remove the average
background level from the ten-second data sets. The azimuth and elevation
angles were then reconstructed for each data set in the ensemble following the
method presented in Section~\ref{dirveccalc}. The mean
reconstructed azimuth and elevation angles from the ensembles, and their
corresponding root mean square deviation, are plotted in
Figure~\ref{fig:dirrec}.
The two plots in Figure~\ref{fig:dirrec} show the
  ability of the ARDUO to reconstruct the direction vector to a 3.2~MBq source
  located 1~to~3~m away 
  using ten seconds of data.

  Figure~\ref{fig:az} shows the reconstructed azimuth angles versus the true
  azimuth angle at elevation angles ranging from -75$^\circ$ 
  to 75$^\circ$. A one-to-one line is plotted
  to compare the true and reconstructed azimuth angles. The reconstructed azimuth
  angles correspond well with the one-to-one line, indicating that the ARDUO
  performs well at reconstructing the azimuth angle for a range of elevation
  angles. Azimuth angle reconstructions for elevation angles above +75$^\circ$ and below -75$^\circ$ were
  not explored as the azimuthal phase space is significantly truncated and vanishing at the poles ($\pm90^\circ$). \par
  Figure~\ref{fig:el} shows the reconstructed elevation
  angles compared with the true elevation angles at azimuth angles ranging
  from 0$^\circ$ to 330$^\circ$. The pink diamonds are results without the
  Responder airframe present. The pink shaded curve
  represents the approximate known system response as obtained from earlier Monte
  Carlo design studies \cite{sinclairARDUOdesign}. Due to limitations in the geometry of the crystals and
  their arrangement, little elevation information is available for points at
  true elevation angles near the horizon. At positive (up) true elevation
  angles with certain azimuth angles, the airframe appears to have an effect on the
  elevation reconstruction. The measurement at a true elevation angle of 15$^\circ$
  and true azimuth angle of 180$^\circ$ appears to be affected by the mass of the
  tailboom. Likewise, the measurement taken at a true
  elevation angle of 50$^\circ$ and a true azimuth angle of 315$^\circ$ appears to be
  affected by the mass of the batteries. As the elevation angle approaches
  vertical, the reconstructed elevation points generally lie near the known
  system response curve.

  \subsection{Two Point Sources}
  Following the laboratory tests, the in-flight performance of the ARDUO
  was examined during a small-scale survey over two 162~MBq point sources of Cs\-/137. 
  The sources were arranged with approximately 27~m separation in a relatively flat survey area of
  approximately 100~m~x~50~m. The source locations were surveyed using a Leica
  Geosystems model GNSS GS15~\cite{Leica}
  to an approximate accuracy of 2~cm~\cite{Al-Khoubbi_2018}. The nominal flight parameters for the ARDUO and Responder system were
  an altitude of 10~m, a speed of 2~m/s and line spacing of 4~m.\par
\begin{figure}[!htb]
    \centering
    \includegraphics[width=0.7\columnwidth]{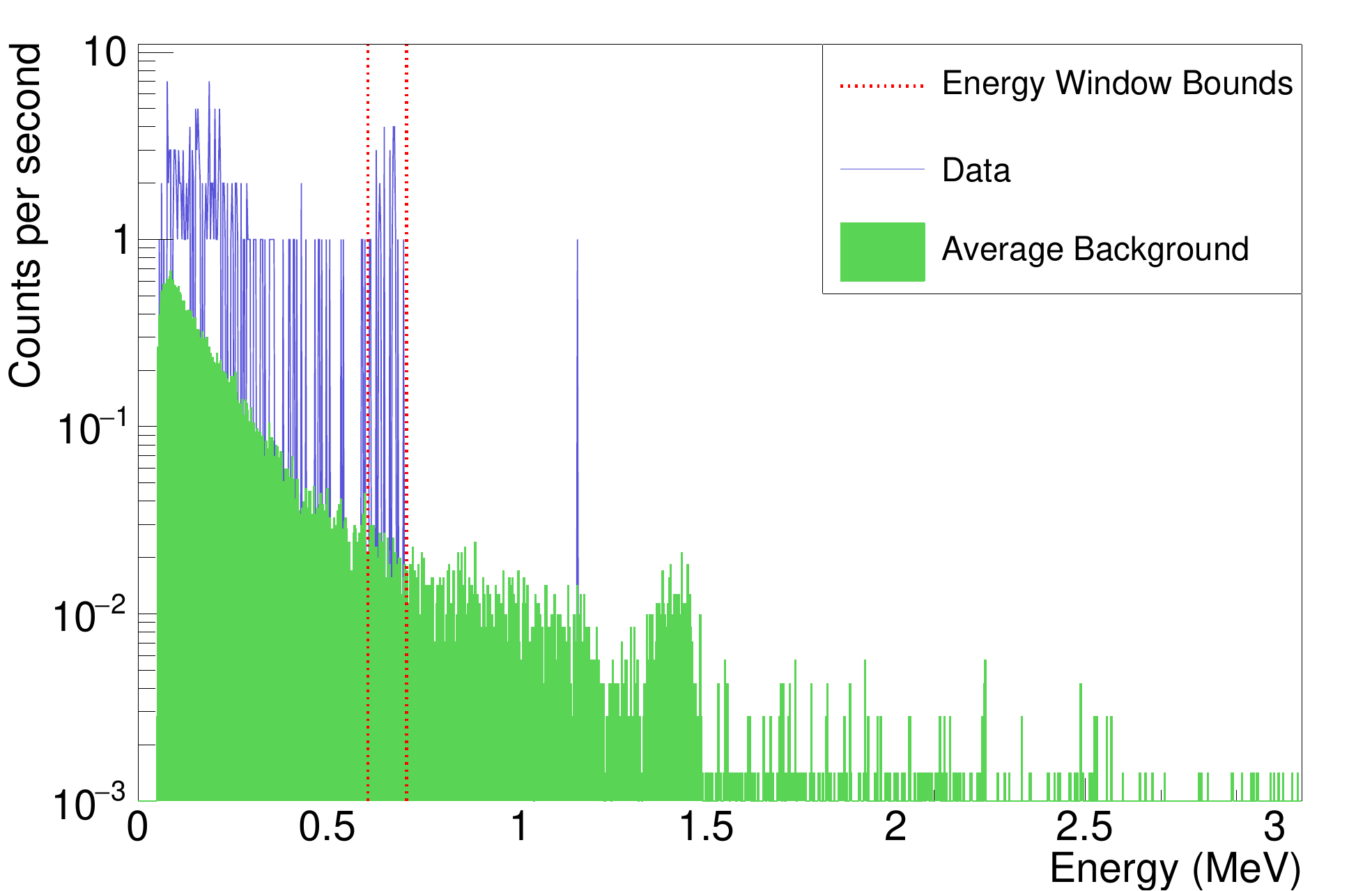}
    \caption{Energy spectra from crystal~1 of the ARDUO when
      conducting a survey over two point sources of Cs\-/137. The
      one-second raw data spectrum is shown by the blue line. The average
      background energy spectrum for the survey
      area is shown by the green-filled area. The energy window bounds used for the photopeak
      analysis are shown by the red dashed lines. The spectrum shows the
      photopeak at 0.662 MeV within the energy window bounds as well as the
low energy deposits from down-scattered Cs-137 and natural background emissions.}
    \label{fig:cs137spec}
  \end{figure}

  Figure~\ref{fig:cs137spec} shows an example one-second spectrum from crystal~1 of
  the ARDUO when flying over the sources of Cs\-/137.
The energy window
  for the count-rate analysis is 100~keV wide (approximately three
  standard deviations) and centered on the
  0.662~MeV photopeak. The spectra from the other seven crystals were also collected,
  and are similar to the example shown. A background survey of the area to determine the contribution to the gamma count rate
  from natural geological sources was conducted on the same
  day prior to source placement, with the same nominal flight parameters. The average background count
  rate is also shown in Fig~\ref{fig:cs137spec}, by the solid green histogram.

\begin{figure}[!htb]
      \centering
    \begin{subfigure}[t]{0.64\columnwidth}
      \centering
      \includegraphics[width=\columnwidth]{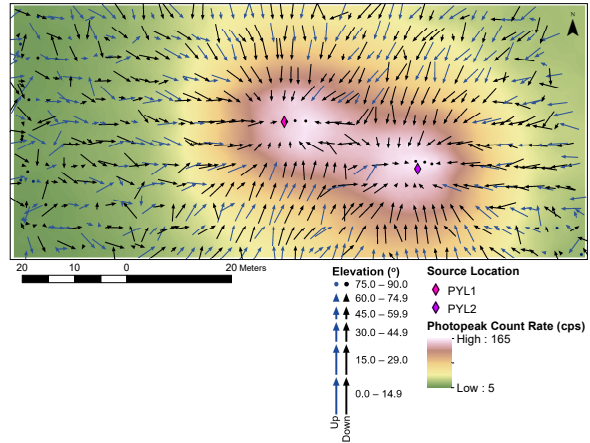}
      \caption{Count rate map.}
      \label{fig:2cs137counts}
    \end{subfigure}
    \begin{subfigure}[t]{0.64\columnwidth}
      \centering
      \includegraphics[width=\columnwidth]{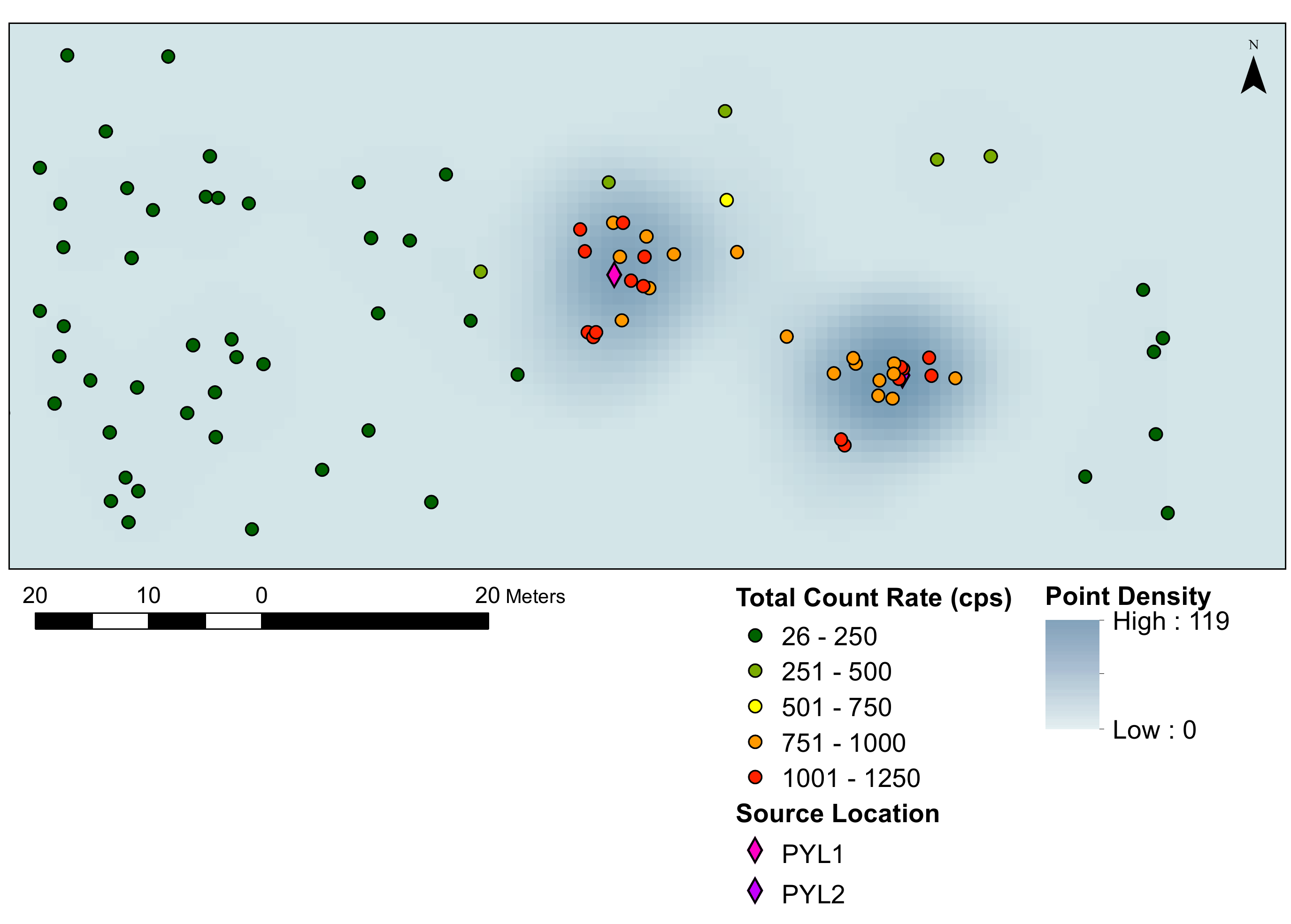}
      \caption{Point density map.}
      \label{fig:2cs137proj}
    \end{subfigure}
    \caption{Maps from the survey over two Cs\-/137
      sources located within 27~m of each other. The true source locations
      are indicated by the pink and purple diamonds and have 2~cm
      uncertainty. a)~is the map of the count rate within an energy window 
      100~keV wide centred on the 0.662~MeV photopeak. White and red indicate a high count rate. The
        direction vectors are overlaid such that shorter arrows are closer to vertical. Blue direction vectors indicate an
        upwards elevation angle, black direction vectors indicate a downwards
        elevation angle.  
b)~is the point density map of the direction vector intersection
      points for those measurements with elevation angle $< 40^\circ$. Darker blue areas correspond to higher point densities. The
      intersection points are coloured according to the count rate
      corresponding to that measurement.}
    \label{fig:2cs137}
  \end{figure}

  The flight data were processed according to the procedure described in
  Section~\ref{processing}.
  The two maps resulting from
  the survey and the post-acquisition data analysis
  can be seen in Figure~\ref{fig:2cs137}.

  Figure~\ref{fig:2cs137counts} shows the count-rate map with the overlaid direction
  vectors and source locations obtained following the methods presented in
Section~\ref{processing}. A high count rate in the Cs\-/137 peak is shown in white and
  red, and a low count rate in the Cs\-/137 peak is shown in green. 
  The true source locations are indicated by the pink and purple diamonds.
  The spatial distribution of the
  count rate shows an elongated anomaly extending diagonally across the
  survey area. The count rate contours outline the general location of the radioactive 
  point sources, however, the result could be interpreted as indicating the presence of a distributed source,
  rather than two point sources.\par
  The reconstructed direction vectors are also shown, overlaid on the count-rate contours in
  Figure~\ref{fig:2cs137counts}. The location of the data point is at the head of the direction
  vector arrow with the length of the vector representing
  its elevation angle and blue arrows pointing up and black arrows pointing
  down.  The direction vectors tend to point towards two
  separate locations on the map, corresponding to the two ends of the elongated anomaly identified using the count-rate contours. Some reconstructed direction vectors
  located between the two true source locations point away from the middle of
  the count-rate anomaly, and toward the two actual source locations near the ends of the anomaly.
  Thus, by examining the direction vectors, it can be surmised that 
  the count-rate distribution is being caused by two separate
  sources. This indicates that the direction vectors are generally providing useful
  information, pointing in the direction of the two true source locations.  
  Of particular benefit during real-time data acquisition, direction vectors
  at the closest approach to the two individual sources tend to point
  vertically downward.

  Figure~\ref{fig:2cs137proj} shows the results from the direction vector
  projection and point density analysis. Using the altitude from the autopilot, the points of intersection of the projected downward-pointing direction
  vectors were calculated and are shown where they intersect the earth's surface.
  The intersection points are weighted by the number of energy
  deposits in the window 0.1~MeV to 3.0~MeV at the measurement point to which they correspond and this weighting is indicated by the colour of the intersection point markers.
  The resulting blue point density contours were calculated using the
  algorithm described in Section~\ref{processing}, with higher densities represented by a darker blue colour.
  The point density contours show that there are two high
  density anomalies within the survey area. This indicates that there are two
  sources in the survey area. 
  The high density anomaly locations nearly centered on the
  true source locations, indicated by the pink and purple diamonds.
  Thus, post-acquisition processing of the
  direction vectors improves the localization of the two sources.

  \subsection{Distributed Source}
  A large-scale distributed source
  survey was conducted at Defence Research and Development
  Canada (DRDC) in Suffield, Alberta. An unmanned ground vehicle (UGV) outfitted
  with an agricultural sprayer distributed an 'L'-shaped pattern of La\-/140
  with dimensions of 120~m~x~20~m for the long arm trending northeast-southwest, and 80~m~x~10~m for the short
  arm trending northwest-southeast. The activity was intended to be approximately uniform at
  10~MBq/m$^2$ within the spray area. The survey location was relatively
  flat. Additional details about the
  experimental setup can be found in \citet{beckman2018}.\par

  Once the La\-/140 was
  distributed, the ARDUO and Responder system flew over the source with a number of different flight parameters.
  The results from only one survey over the source are presented here. The nominal flight
  parameters for the survey were an
  altitude of 10~m, a speed of 10~m/s, and a line spacing of 10~m.

\begin{figure}[!htb]
    \centering
    \includegraphics[width=0.7\columnwidth]{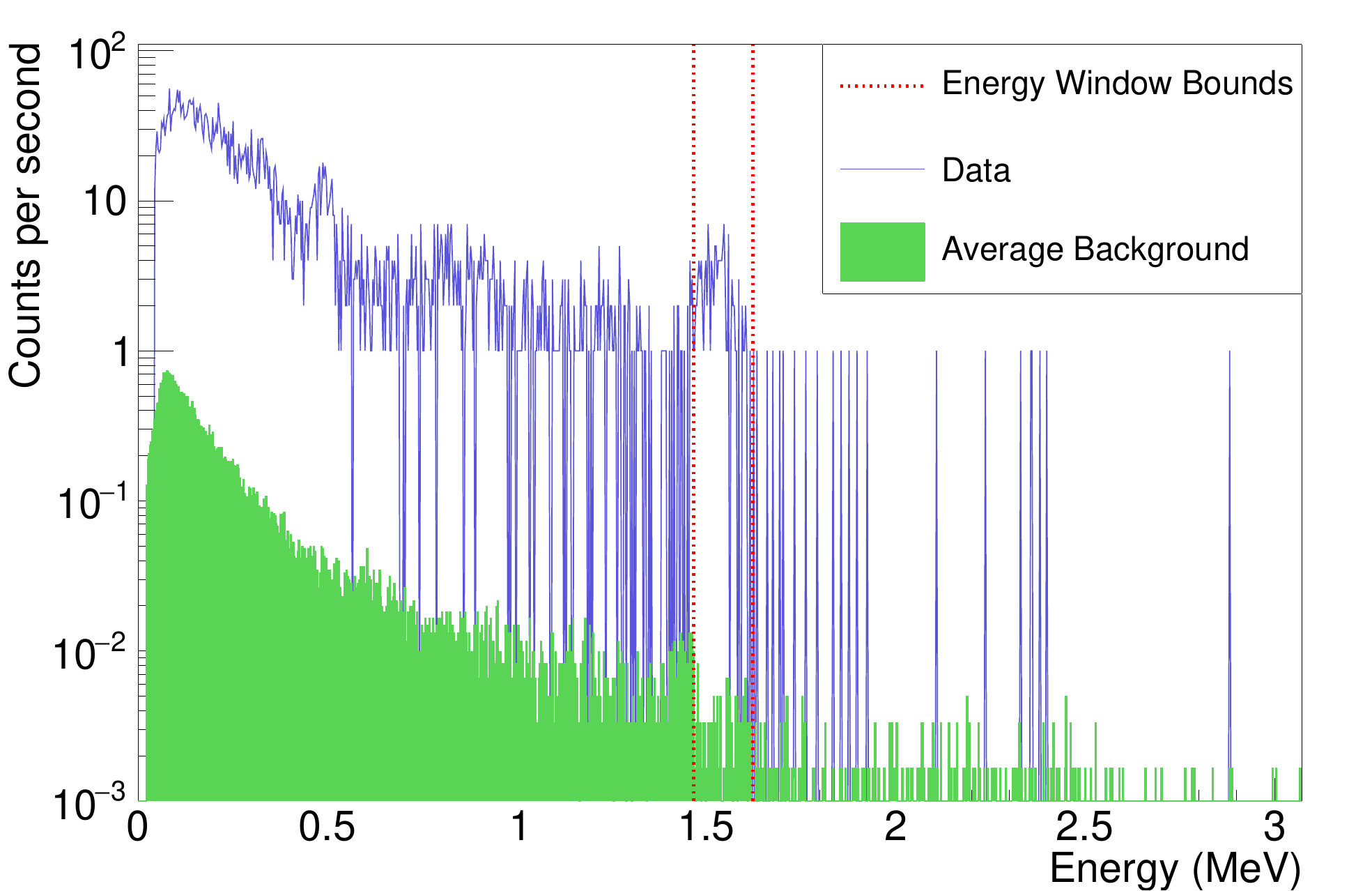}
    \caption{Energy spectra from crystal 1 of the ARDUO while
      flying over the source of La\-/140. The
      one-second raw data spectrum is shown by the blue line. The average
      background energy spectrum for the survey
      area is shown by the green-filled area. The energy window bounds used for the photopeak
      analysis are shown by the red dashed lines. The spectrum shows the
      photopeak at 1.596~MeV within the energy window bounds as well as lower
energy La\-/140 emissions and deposits from downscattered gamma-rays of both
La\-/140 and natural background.}
    \label{fig:La140_spectrum}
  \end{figure}

  La\-/140 is an isotope with multiple gamma-ray energy emissions.  A typical
  one-second spectrum accumulated during the flight over the source is
  shown in Figure~\ref{fig:La140_spectrum}. 
The counts falling into a three standard deviation energy
  window 150~keV wide and centered around La\-/140's highest
  energy photopeak at 1.596~MeV were used in the analysis.
  A background flight over the survey area was conducted prior to the distribution of La\-/140. The average background energy spectrum is overlaid on the signal spectrum in 
  Figure~\ref{fig:La140_spectrum}.

\begin{figure}[!htb]
   \centering
    \begin{subfigure}[t]{0.55\columnwidth}
      \centering
      \includegraphics[width=\columnwidth]{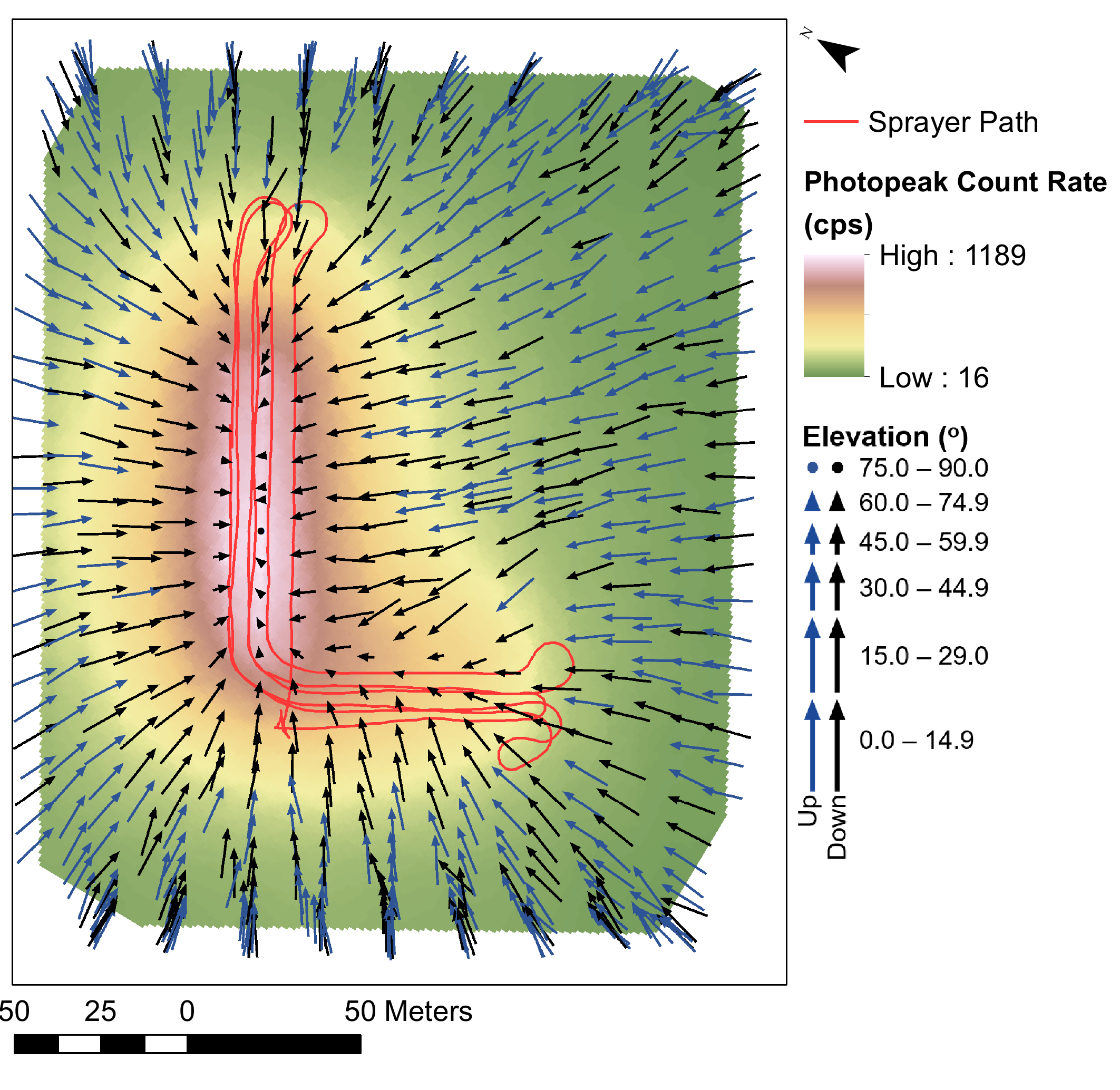}
      \caption{Count rate map.}
      \label{fig:suffieldcounts}
    \end{subfigure}
    \begin{subfigure}[t]{0.55\columnwidth}
      \centering
      \includegraphics[width=\columnwidth]{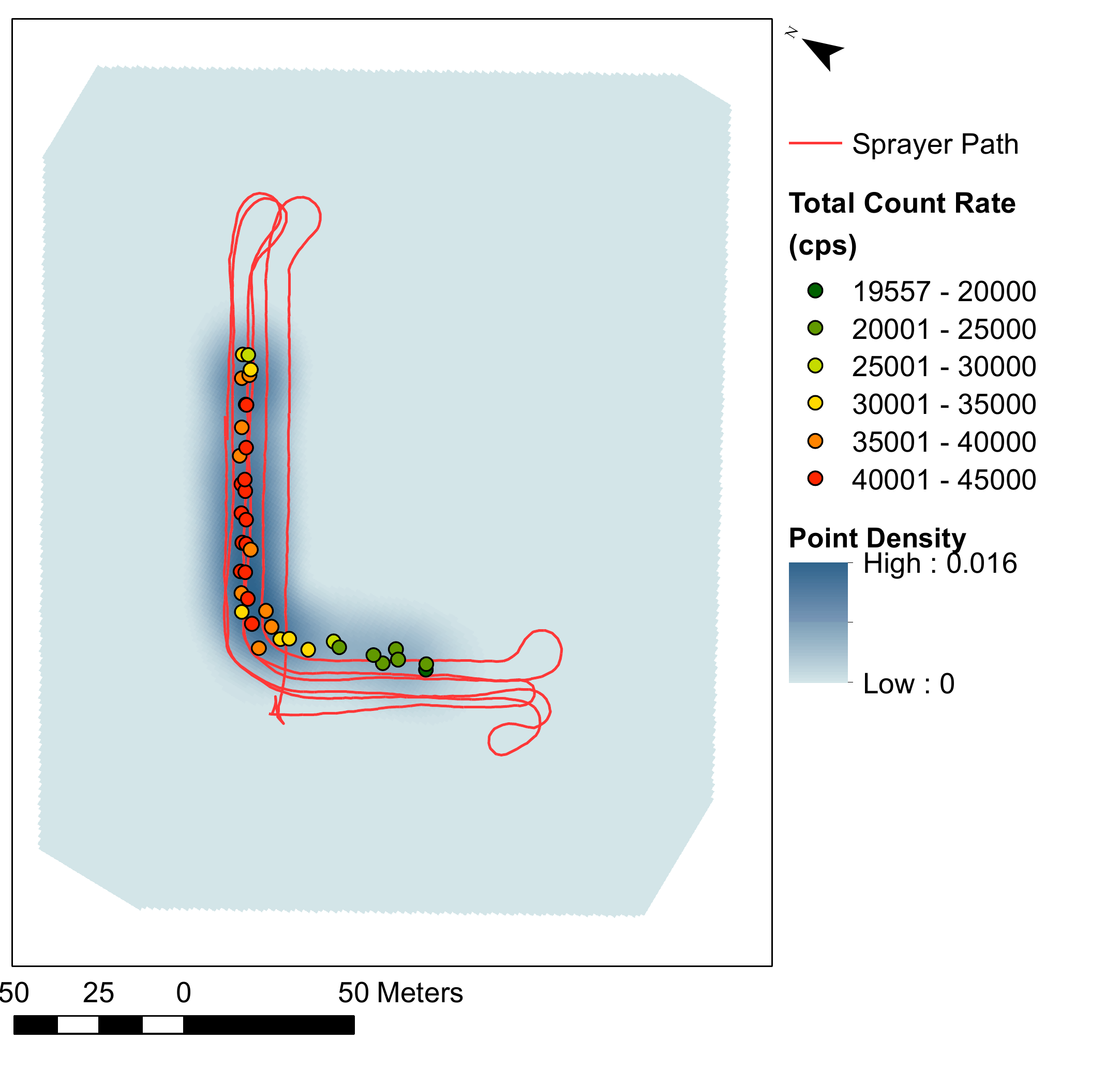}
      \caption{Point density map.}
      \label{fig:suffieldproj}
    \end{subfigure}
    \caption{The maps from the distributed La\-/140 survey from DRDC
      Suffield. The path of the radiation spraying UGV is shown by
      the red lines as an approximation of the ground truth. a)~is the map of
      the count rate within the 150~keV-wide energy window about the 1.596~MeV
      photopeak. The
      direction vectors are overlaid such that shorter arrows are closer to
      vertical. Blue direction vectors indicate an
      upwards elevation angle, black direction vectors indicate a downwards
      elevation angle.  b)~is the point density map of the direction vector intersection
      points for those measurements with elevation angle $< 40^\circ$. Darker blue areas correspond to higher point densities. The
      intersection points are coloured according to the count rate
      corresponding to that measurement.}
    \label{fig:suffield}
  \end{figure}

  Post acquisition, the data were analyzed in the same way as the point source data, following 
  the methods presented in Section~\ref{processing}.  The results of the post-acquisition analysis are shown in
  Figure~\ref{fig:suffield}.
  Similar to the map produced for the small-scale point-source survey,
  Figure~\ref{fig:suffieldcounts}  shows the count-rate map with the overlaid direction
  vectors. The path of the UGV sprayer is shown by the red lines. This is
  an approximation of the ground truth for where the distributed source is
  located. The spatial distribution of the count rate
  indicates that there is a high linear anomaly trending
  northeast-southwest. It also shows a smaller linear anomaly trending
  northwest-southeast, perpendicular to the higher anomaly. The spatial
  distribution of the count rate is thus consistent with the
  known location of the sprayer path, however the count rate contours extend
  more broadly than the known path. \par
  The reconstructed direction vectors are also shown in Figure~\ref{fig:suffieldcounts} overlaid on the count-rate contours.
  In contrast to the small-scale point-source survey, the direction vectors point along the lengths of both
  linear anomalies, therefore indicating that the source of the anomalies
  is distributed along two linear paths.
  The direction vectors show good correlation to the sprayer
  path and indicate that the source may be more narrowly distributed than the
  count-rate contours suggest. \par
  Figure~\ref{fig:suffieldproj} shows the intersection points with the flat earth's surface of the projected downward-pointing direction vectors. The colour of each
  intersection point represents its weighting from the number of measured
  energy deposits in the range 0.1~MeV to 3.0~MeV.
  Similar to Figure~\ref{fig:2cs137proj}, the point densities are contoured with higher point densities shown
  in darker blue. Again, the sprayer path is overlaid as shown by the red
  lines. The map shows a high point density linear area trending northeast-southwest, and a less dense linear
  area trending northwest-southeast. The two features appear perpendicular to
  each other and correlate well with the sprayer path. These features are
  similar to the anomalies found in the count-rate map
  (Figure~\ref{fig:suffieldcounts}), but the features in the point density map
  have a reduced spatial extent. The area of each feature in the point density
  follows more closely with the extent of the sprayer path. This indicates that the direction projection
  processing method can improve source localization.

  \section{Summary and Conclusions}
  The ARDUO is a novel detector of gamma radiation that is mounted on a
  UAV. The ARDUO meaures gamma energy spectra as a function of position and also determines the direction of radioactivity using its self-shielding properties.
  The ARDUO and UAV system can conduct aerial radiometric surveys and display the data as well
  as directional information in real time, allowing for in-flight source
  locating. In post-processing, the directional information can be used to
  improve source locating.
  Mapping of the direction vectors appears to provide a superior ablility to distinguish
  multiple point sources and 
  spatially restrict distributed sources, improving on mapping the gamma count
  rates alone.
  A point density map may be produced by projecting the data points from the UAV altitude
  to the ground and looking at the intersection points.
  The point density contours from the projected direction vectors more closely delineate the 
  spatial distribution of the source than the simple count-rate contours.
  The direction vector projection method has been demonstrated 
  by distinguishing two point sources with a 27~m separation, from an altitude
  approximately half this amount.
  The same method has also demonstrated to better constrain the spatial extent of a large distributed source.

  \section{Acknowledgments}
  This work was supported by Canada's Centre for Security Science as a
  part of CSSP-2016-TI-2290. We would like to thank Defence Research and Development
  Canada Suffield for their collaboration on the distributed source survey. This paper is NRCan Contribution number 20180165.
  
  \bibliography{references_2}  

\end{document}